\documentclass[twocolumn,showpacs,amsmath,amssymb,aps,pra,footinbib,floatfix,nobalancelastpage,superscriptaddress,10pt]{revtex4-1}
\usepackage{graphicx}
\usepackage{dcolumn}
\usepackage{bm}
\usepackage{helvet}
\usepackage{amssymb}
\usepackage{amsmath}
\usepackage{amsfonts}
\usepackage{color}
\newcommand{\bra}[1]{\langle#1|}
\newcommand{\ket}[1]{|#1\rangle}
\newcommand{\beq}{\begin{equation}}
\newcommand{\eeq}{\end{equation}}

\newcommand{\sandwich}[3]{\langle#1|#2|#3\rangle}
\newcommand{\eva}[1]{\langle#1\rangle}

\newcommand{\comm}[2]{\left[#1,#2\right]}
\newcommand{\WW}{\Omega}
\newcommand{\ww}{\omega}
\newcommand{\oo}[1]{\mathcal{O}(g^{#1})}
\newcommand{\BK}{{\bm{k}}}
\newcommand{\BKprime}{{\bm{k'}}}

\begin{document}

  \title{Stroboscopic prethermalization in weakly interacting periodically driven systems}
  \author{Elena Canovi}
  \affiliation{Max Planck Research Department for Structural Dynamics, University of Hamburg-CFEL, Hamburg, Germany}
  
  \author{Marcus Kollar}
  \affiliation{Institut f\"ur Physik, Theoretische Physik III, 
    Center for Electronic Correlations and Magnetism, University of Augsburg, Augsburg, Germany}
  
  \author{Martin Eckstein}
  \affiliation{Max Planck Research Department for Structural Dynamics, University of Hamburg-CFEL, Hamburg, Germany}

\date{\today}  

  \begin{abstract}
    Time-periodic driving provides a promising route to engineer
    non-trivial states in quantum many-body systems. However, while it
    has been shown that the dynamics of {\it integrable, non-interacting} systems can
    synchronize with the driving into a non-trivial periodic motion,
    generic {\it non-integrable} systems are expected to heat up until
    they display a trivial infinite-temperature behavior. In this
    paper we show that a quasi-periodic time evolution over many
    periods can also emerge in 
	weakly interacting 
    systems, 
     with a clear separation of the timescales for
    synchronization and the eventual approach of the
    infinite-temperature state. This behavior is the analogue of
    prethermalization in quenched systems. The synchronized state can
    be described using a macroscopic number of approximate constants
    of motion. We corroborate these findings with numerical
    simulations for the driven Hubbard model.
  \end{abstract}

  
  \maketitle

\section{Introduction}
Experiments with ultra-cold atomic gases in optical lattices and ultra-fast spectroscopy 
nowadays allow to address the dynamics of quantum many-particle systems
  out of equilibrium.
  A particularly important role in this context is played by
  periodically driven systems~\cite{Shirley1965,Sambe1973,Breuer_JPF90, GrifoniHanggi1998}. 
  Periodic driving can stabilize novel
  states both in cold atoms and in condensed matter, including
  topologically nontrivial states~\cite{Oka_PRB09, Kitagawa_PRB11,
    Iadecola_PRL13, KitagawaPRB10,Lindner_NP11, Wang_SCI13} or complex
  phases such as superconductivity~\cite{Fausti11,Hu2014}. It can be
  used to engineer artificial gauge fields in cold
  atoms~\cite{Goldman_PRX14} and emergent many-body interactions such
  as magnetic exchange interactions in
  solids~\cite{Mentink2015,Mikhaylovskiy2014, Itin_pp14}, or to transiently
  modify lattice structures through anharmonic
  couplings~\cite{Foerst2011}.

  An important question is thus the theoretical understanding of the long-time dynamics of periodically driven 
  systems.
  The approach to a steady state has been investigated intensively for
  the relaxation of isolated systems after a sudden perturbation, both
  experimentally and theoretically~\cite{Polkovnikov2011RMP,
    Kinoshita2006, Trotzky2012}. When a generic non-integrable
  many-body system is left to evolve with a time-independent
  Hamiltonian, it is believed to eventually relax to a thermal
  equilibrium state, unless it is in a many-body localized phase~\cite{Altshuler_PRL97,Gornyi_PRL05,Basko_AP06}. 
  If the system is integrable, on the other hand,
  the steady state is often described by a generalized Gibbs ensemble
  (GGE)~\cite{Rigol2007, Langen2014}, which keeps track of a
  macroscopic number of constants of motion. When integrability is
  only slightly broken, the system can display dynamics on separate
  timescales, such that observables rapidly {\em prethermalize} to a
  quasi-steady nonequilibrium state which can be understood by a GGE
  based on perturbatively constructed constants of
  motion~\cite{Kollar_PRB11}, before thermalizing on much longer time
  scales~\cite{Berges2004a, Moeckel2008a, Gring2012}. 

  Integrability turns out to be a crucial factor also for periodically
  driven systems. Their dynamics can synchronize with the
  driving~\cite{Russomanno_PRL12} and display a non-trivial periodic
  time evolution at long times.  A way to understand this is to show
  that the time evolution over one period $T$ commutes with an
  infinite number of operators $\mathcal{I}_\lambda$, which are thus
  conserved at stroboscopic times (i.e. integer multiples of the
  period). Having a fixed expectation value of all
  $\mathcal{I}_\lambda$ at stroboscopic times, one can construct a
  statistical ensemble to describe the long-time behavior of the
  system (the periodic Gibbs ensemble), which has been analytically
  and numerically shown to give correct predictions for hard-core
  bosons~\cite{Lazarides_PRL14}.

  In contrast to integrable systems (and many-body localized
  states~\cite{Lazarides_PRL15,Abanin_pp14,DAlessio_AP13, Ponte_AP15, Ponte_PRL15, Roy_PRB15}),
   it has been proposed that generic non-integrable systems ``heat up''
  under the effect of driving and display rather trivial infinite
  temperature properties as soon as they settle into a periodic
  motion~\cite{Lazarides_PRE14, DAlessio_PRX14,Eckardt_pp15}. One can
  formulate this statement in terms of the Floquet eigenstates (the
  exact solutions of the Schr\"odinger equation with a periodic
  evolution of all observables~\cite{Dittrich1998,
    GrifoniHanggi1998}), stating that each individual Floquet state
  displays infinite temperature properties. This conjecture relies on
  a breakdown of the perturbative expansion of Floquet eigenstates at
  some order because of unavoidable resonances between transitions in
  the many-body spectrum with multiples of the driving frequency. 
  A common approach to avoid this problem is to construct effective 
  Floquet Hamiltonians from a high-frequency expansion \cite{DAlessio_AP13,Goldman_PRX14}. 
  In this work we show that a quasi-periodic state can also emerge 
  in weakly interacting systems, 
  provided 
  that linear absorption can be avoided:
  the
  stroboscopic time evolution is constrained by approximately
  conserved constants of motion
  $\tilde{\mathcal{I}}_\lambda$. Analogously to prethermalization in
  weakly interacting systems after a sudden
  perturbation~\cite{Berges2004a, Moeckel2008a, Kollar_PRB11}, the
  system rapidly synchronizes with the driving and remains periodic
  over a large number of periods $m$ such that $g^{-1} \gg mT \gg
  J^{-1}$, where $g$ controls the strength of the interaction, 
  $T$ is the period
  of the driving and $J$ is coupling of the integrable Hamiltonian,
  e.g., the bandwidth of the kinetic energy term; we set $\hbar$ $=$
  $1$ throughout.  The quasi-periodic state can be described as a
  periodic Gibbs ensemble based on the $\tilde{\mathcal{I}}_\lambda$,
  i.e., stroboscopic prethermalization gives access to quasi-periodic
  states which are entirely different from the infinite temperature
  final states.

  This paper is organized as follows: in Sec.~\ref{sec:general} we introduce the general formalism: first, we rotate the weakly interacting Hamiltonian in such a way
  that it commutes with the integrals of motion of the noninteracting part (Sec.~\ref{ssec:rotation}), then in Sec.~\ref{ssec:integrals} we identify the approximate integrals of motion and finally in Sec.~\ref{ssec:obs} study the time evolution 
  of the observables. In Sec.~\ref{sec:PGE} we discuss the relation with the periodic Gibbs ensemble~\cite{Lazarides_PRL14} and in Sec.~\ref{sec:floquet} the relation with
  the Floquet theory of periodically driven systems. In Sec.~\ref{sec:hubbard} we
  specialize the results of Sec.~\ref{sec:general} to the Hubbard model, first presenting our analytical results (Sec.~\ref{ssec:analytical}) and then comparing them to the numerical findings (Sec.~\ref{ssec:numerical}) obtained with dynamical mean-field theory. In Sec.~\ref{sec:conclusions} we draw our conclusions.

  \section{General formalism}\label{sec:general}
  In the following we consider an
  integrable, noninteracting 
  system perturbed by 
  a weak periodic interaction. 
  The general Hamiltonian is given by
  \beq\label{eq:ham}
  H(t)=H_{0}+ g H_{\rm int}(t),
  \eeq
  where the integrable part 
  \beq
  H_0=\sum_{\lambda}\epsilon_{\lambda}\hat {\mathcal{I}}_{\lambda}\;,
  \eeq
  can be written as a sum of constants of motion (e.g., momentum
  occupations for independent particles on a lattice), and the small
  parameter $g$ controls the strength of the interaction $H_{\rm
    int}(t)$ which is periodic with period $T$ and frequency
  $\WW=2\pi/T$. To study the time evolution at stroboscopic times
  $t_m=mT$ ($m$ integer), we extend the approach of
  Refs.~\cite{Kollar_PRB11, Moeckel2008a} to periodically driven
  systems, and determine a time-periodic unitary transformation $R(t)$
  such that the Hamiltonian $H_{\rm eff}(t)$ in the rotated frame
  commutes with the constants of motion $\hat{\mathcal I}_\lambda$
  {\it at any time} up to corrections or order $\oo{3}$
  \footnote{Unitary transformations of the periodic Hamiltonian $H(t)$ are also used in Ref.~\onlinecite{Abanin_pp14}, but in that case with the aim of finding a many-body localized cycle Hamiltonian $e^{i H_{\rm cycle}T}\equiv \mathcal{T} e^{-i\int_{0}^t dt'\,H(t')}$, i.e. a Hamiltonian with a set of true integrals of motion.}. 
  If
  $\ket{\tilde{\psi}(t)}=R(t)\ket{\psi(t)}$ 
  is 
  the transformed
  wave function, the Hamiltonian $H_{\rm eff}$ which dictates the
  evolution in the rotated frame via $i\partial_t
  \ket{\tilde{\psi}(t)} = H_{\rm eff} (t)\ket{\tilde{\psi}(t)}$ is
  given by
  \beq\label{eq:Heff0}
  \begin{split}
    H_{\rm eff}(t)&= R(t)H(t)R(t)^\dagger-iR(t)\dot R(t)^{\dagger}\;.
  \end{split}
  \eeq
Since $R(t)$ is assumed to be unitary, we make the ansatz $R(t)\equiv e^{S(t)}$, with an anti-hermitian
  operator $S(t)$.

  In the next sections, we first analytically find the transformation $S(t)$ (Sec.~\ref{ssec:rotation}), 
  then identify the approximate integrals of motion (Sec.~\ref{ssec:integrals}) and  finally in Sec.~\ref{ssec:obs} we  study the time evolution of the expectation values of observables and the dependence of their long-time behavior on the frequency $\Omega$.
 \subsection{The transformation $S(t)$}\label{ssec:rotation}
    We now show in
  detail how to rotate the Hamiltonian Eq.~\eqref{eq:ham} with a
  transformation $R(t)$ $=$ $e^{S(t)}$ such that the Hamiltonian~\eqref{eq:Heff0}
  is (i) periodic and (ii) diagonal in the operators that diagonalize
  $H_0$.  To implement condition (i), we first expand to second order
  in $g$ and then write in Fourier series both the effective
  Hamiltonian
  \begin{equation}\label{eq:HeffFourier}
  \begin{split}
    H_{\rm eff}(t)&=H_{\rm eff}^{(0)} (t)+g H_{\rm eff}^{(1)} (t)+g^{2} H_{\rm eff}^{(2)} (t)\\
    &=\sum_{n} e^{-in\WW t}\left[H_{\rm eff,n}^{(0)}+g H_{\rm eff,n}^{(1)} +g^{2}H_{\rm eff,n}^{(2)}\right]
  \end{split}
  \end{equation}
  and the anti-hermitian operator
  \begin{equation} 
  \begin{split}
    S(t)&=gS^{(1)}(t)+\frac{g^2}{2}S^{(2)}(t)+\mathcal{O}(g^3)\\
    &=\sum_{n} e^{-in\WW t} [g S^{(1)}_{n}+\frac{g^{2}}{2}S^{(2)}_{n}]+\mathcal{O}(g^3)\;,
  \end{split}
  \end{equation}
  with $H_{{\rm eff},n}=H_{{\rm eff},-n}^{\dagger}$  and $S_{n}=-S_{-n}^{\dagger}$.
  Combining Eqs.~\eqref{eq:Heff0} and~\eqref{eq:HeffFourier} we find:
  \begin{equation}\label{eq:Hefford}
  \begin{split}
    &H_{\rm eff}(t)=H_0+g\left(H_{\rm int}(t)+[S^{(1)}(t),H_0]+i\frac{d}{dt}S^{(1)}(t)\right)\\
    &+g^2\left(\frac{1}{2}[S^{(2)}(t),H_0]+[S^{(1)}(t),H_{\rm int}(t)]+\right.\\
    &\frac{1}{2}[S^{(1)}(t),[S^{(1)}(t),H_0]]+\frac{i}{2}\frac{d}{dt}S^{(2)}(t)+\\
    &\left.-\frac{i}{2}(\dot S^{(1)}(t) S^{(1)}(t)-S^{(1)}(t)\dot S^{(1)}(t))\right)+\mathcal O(g^3)\;.
  \end{split}
  \end{equation}
  To ensure condition (ii), we require that 
  \begin{equation}\label{eq:commdiag}
  [H_{\rm eff,n}^{(X)},\hat {\mathcal{I}}_{\lambda}]=0,
  \end{equation}
  for any Fourier component, perturbative order and constant of
  motion, labeled by $n$, $X$ and $\lambda$ respectively. As in
  Ref.~\onlinecite{Kollar_PRB11}, we employ the basis $\hat
  {\mathcal{I}}_{\lambda}\ket{\boldsymbol\alpha}=\alpha_\lambda\ket{\boldsymbol\alpha}$. 
  We assume that the energies $\epsilon_{\lambda}$ are
  incommensurate, so that the eigenenergies of $H_0$,
  i.e. $E_{\boldsymbol{\alpha}}=\sum_{\lambda}\epsilon_{\lambda}\alpha_{\lambda}$,
  are nondegenerate.
  For an extensive lattice model this can be achieved, e.g., by using sufficiently irregular boundaries.
   
   After some lengthy but otherwise straightforward
  algebra, we can find $H_{\rm eff}(t)$ and $S(t)$ by repeatedly
  applying Eq.~\eqref{eq:commdiag} to each perturbative order in
  Eq.~\eqref{eq:Hefford}, so as to reduce the Hamiltonian to the diagonal form
 \beq\label{eq:Hdiagper}
  H_{\rm eff}(t)= H_0 + \sum_\alpha\ket{\alpha}E_{\rm diag,\alpha}(t)\bra{\alpha} +\oo{3}\;.
  \eeq

  In order $g^0$ we have 
  \begin{equation}\label{eq:H00}
  H_{\rm eff,n}^{(0)}=\left\lbrace
    \begin{array}{ll}
      H_0&\qquad{\rm if}\; n=0\\
      0&\qquad{\rm otherwise}\;,
    \end{array}
  \right.
  \end{equation}
  so that $H_{\rm
    eff,0}^{(0)}=\sum_\alpha\ket{\alpha}\bra{\alpha}E^{(0)}_{0,\alpha}$,
  with $E^{(0)}_{0,\alpha}=E_\alpha$.

  To first order in $g$ the Fourier components of $S(t)$ read:
  \begin{equation}\label{eq:S1}
  \sandwich{\beta}{S_n^{(1)}}{\alpha}=\left\lbrace
    \begin{array}{ll}
      \frac{\sandwich{\beta}{{H_{\rm int,n}}}{\alpha}}{E_\beta-E_\alpha-n\WW}&{\rm if}\quad \alpha\neq \beta\\
      0&\qquad{\rm otherwise}
    \end{array}
  \right.
  \end{equation}
  The first-order perturbative correction to $H_{\rm eff}$ is:
  \begin{equation}
  \begin{split}
    H_{\rm eff}^{(1)}(t)&=\sum_\alpha e^{-in\WW t}\ket{\alpha}E^{(1)}_{n,\alpha}\bra{\alpha}
  \end{split}
  \end{equation}
  where 
  \begin{equation}\label{eq:E1}
  E^{(1)}_{n,\alpha}=\sandwich{\alpha}{H_{\rm int,n}}{\alpha}\;.
  \end{equation}

  At order $g^2$  the Fourier components of $S^{(2)}$ are found to be:
  \begin{equation}\label{eq:S2gen}
    \sandwich{\beta}{S^{(2)}_n}{\alpha}=\sum_p \frac{\sandwich{\beta}{\comm{S_p^{(1)}}{H_{{\rm int},n-p}+ H^{(1)}_{{\rm diag},n-p}}}{\alpha}}{E_\beta-E_\alpha-n \Omega}
  \end{equation}
  if $\alpha\neq\beta$ and, as previously, we choose the diagonal elements to be zero. 
 In Eq.~\eqref{eq:S2gen} we have defined:
  $H^{(1)}_{{\rm diag},n}=\sum_{\alpha}\ket{\alpha}E^{(1)}_{n,\alpha}\bra{\alpha}$.
 Finally, the second order term of the effective Hamiltonian reads:
  \begin{equation}
  H^{(2)}_{{\rm eff},n}=\sum_\alpha\ket{\alpha}E^{(2)}_{n,\alpha}\bra{\alpha}\;,
  \end{equation}
  with:
 \begin{equation}
  \begin{split}
    E^{(2)}_{n,\alpha}=&\frac{1}{2}\sum_{\beta\neq \alpha}\sum_p\left[\frac{\sandwich{\alpha}{H_{{\rm int},p}}{\beta}\sandwich{\beta}{H_{{\rm int},n-p}}{\alpha}}{E_\alpha-E_\beta-p\Omega}\right.\\
       &-\left.
       \frac{\sandwich{\alpha}{H_{{\rm int},n-p}}{\beta}\sandwich{\beta}{H_{{\rm int},p}}{\alpha}}{E_\beta-E_\alpha-p\Omega}
       \right]\;.
  \end{split}
  \end{equation}

   \subsection{Approximate integrals of motion}\label{ssec:integrals}
  Under a general unitary transformation, the time propagator
  $U(t,0)$ $=$ $\mathcal{T} e^{-i\int_{0}^t dt'\,H(t')}$ is transformed into
  \beq
  \label{timeevolution}
  U(t,0)
  =
  e^{-S(t)} \,\tilde U(t,0)\,e^{S(0)},
  \eeq
  with $\tilde U(t,0)$ $=$ $\mathcal{T} e^{-i\int_{0}^t dt'\,H_{\rm
      eff}(t')}$.  Because $S(t)$ is periodic, the time evolution at
  stroboscopic times is thus unitarily equivalent to the time
  evolution with the diagonal Hamiltonian \eqref{eq:Hdiagper},
  $U(t_m,0) = e^{-S(0)} e^{-i\int_{0}^{t_m} dt\,H_{\rm eff}(t)}
  e^{S(0)} + t_m\oo{3}$. This implies that the quantities
  \beq
  \label{scom}
  \tilde 
  {\mathcal I}_\lambda
  =
  e^{-S(0)} 
  \hat{\mathcal I}_\lambda
  e^{S(0)} 
  \eeq
  are approximately conserved 
  under the evolution over multiple periods $T$,
  i.e., 
  $\langle 
  \tilde 
  {\mathcal I}_\lambda(t_m)
  \rangle
  =
  \langle 
  \tilde 
  {\mathcal I}_\lambda(0)
  \rangle
  +t_m\oo{3}$. 
  For the example of a weakly interacting Hubbard model studied below, the original
  constants of motion are momentum occupations $n_{\BK}$ of
  independent particles, while the constants of motion of the
  stroboscopic time evolution correspond to quasiparticle modes.
   \subsection{Expectation value of observables}\label{ssec:obs}
  We examine 
  the synchronization  of these modes
  in terms of the time evolution 
  \beq
  \label{At}
  \eva{A}_{t}\equiv\sandwich{\psi(0)}{U^{\dagger}(t,0)AU(t,0)}{\psi(0)}\;
  \eeq
  of an observable $\hat A$ which is a function of the original
  constants of motion $\mathcal{I}_{\lambda}$ (having in mind, e.g., a
  measurement of momentum occupations $n_{\BK}$ or higher-order
  momentum correlation functions $n_{\BK}n_{\BKprime}$), assuming that
  the system is in an eigenstate $\ket{\psi(0)}\equiv \ket{\alpha}$ of
  $H_0$ before the driving is switched on. Inserting
  Eq.~\eqref{timeevolution} into \eqref{At}, expanding the operators
  $e^{S(0)}$ and $e^{S(t)}$ in powers of $g$, and using the fact that
  $[A,\tilde U(t,0)]=t\oo{3}$ (because $A$ commutes with all
  $\mathcal{I}_\lambda$), we obtain
  \beq\label{eq:b2periodic-1}
  \eva{A}_t
  =
  -2{\rm Re} \sandwich{\alpha}{S(0)A[S(0)-\bar S(t)]}{\alpha}+t\oo{3},
  \eeq
  with $\bar S(t)\equiv \tilde U^\dagger(t,0)S(t)\tilde U(t,0)\;$.
  For stroboscopic times, with $S(t_m)=S(0)$ determined by
  Eq.~\eqref{eq:S1}, one finds the final result for the perturbative time
  evolution
  \begin{align}\label{eq:expvAmain}
    \!\!\!\eva{A}_{t_m}=\sum_{n,p} \int\limits_{-\infty}^{\infty}\!\!d\ww\,
    \frac{4g^2\sin^2(\frac{\ww t_m}{2}) y_{np}(\ww)}{(\ww - n\WW)(\ww - p\WW)}
    + t_m\oo{3},
  \end{align}
  where $y_{np}(\ww)$ denotes the spectral density
  \beq\label{eq:yfunction-1}
  \begin{split}
    y_{np}(\ww)
    =&\sandwich{\alpha}{H_{{\rm int},-n}A\delta(\ww-H_0 + E_\alpha)H_{{\rm int},-p}}{\alpha}\;.
  \end{split}
  \eeq
  The integral in Eq.~\eqref{eq:expvAmain} gives an accurate
  description of $\eva{A}_{t_m}$ for times $t_m\ll g^{-1}$, where
  relative corrections $t_m\oo{3}$ are small.
  Note that for finite
  $m$ the term $\sin(t_m\ww/2)$ regularizes the singularities at
  $n\ww$.
  Therefore the amount of contributing spectral weight is due
  to the location of $\Omega$ inside or outside the band,
  as discussed below.
  For $g\to0$ there is thus a large time window $g^{-1} \gg
  t_m \gg T $ in which the dynamics is governed by the long time
  asymptotics of the integral. To analyze this, we distinguish two
  different behaviors depending on the frequency $\WW$:

  {\em (i) Fermi golden rule regime:} If there is nonzero spectral
  density $y_{nn}(n\WW)>0$ at an even pole $1/(\ww - n\WW)^2 $, the
  stroboscopic evolution for $m\gg1$ develops a linear asymptotics
  $\eva{A}_{t_m} \sim g^2 t_m \sum_{\beta\neq\alpha }
  \bra{\beta}A\ket{\beta}
  \Gamma_{\alpha\to\beta}$, 
  where $\Gamma_{\alpha\to\beta}$ is the Fermi golden rule excitation
  rate
  \begin{align} 
    \Gamma_{\alpha\to\beta}
    =
    2\pi
    \sum_{n}  |\langle\beta|H_{\rm int,n}|\alpha\rangle|^2 \delta(n\WW -E_\beta+E_\alpha).
  \end{align}
  To see this fact one can consider the contribution to the integral
  \eqref{eq:expvAmain} from a small interval $|\ww-n\WW|\le\epsilon$
  around the pole, in which $y_{nn}(\ww)$ can be approximated by a
  constant $y_{nn}(n\WW)$. With a substitution $x=t_m(\ww-n\WW)$, the
  remaining integral is $t_m\int_{-\epsilon t_m}^{\epsilon t_m}dx
  \,\sin^2(x/2)/ x^2 \sim t_m\pi/2 $. From a similar consideration for
  $n$ $\neq$ $p$ one can obtain the subleading terms.

  {\em (ii) Stroboscopic prethermalization:} Assuming that the
  perturbation involves only a limited number of Fourier components,
  such as for a harmonic perturbation with $H_{{\rm int},n}=0$ for
  $|n|>1$, then the spectral density $y_{nm}(\ww)$ is restricted to a
  finite band $[-W,W]$, depending on the type of excitation, the
  bandwidth of the noninteracting single-particle spectrum, and phase
  space restrictions. If all poles $\ww=n\WW$ lie outside this band,
  the limit $m\to\infty$ integral of \eqref{eq:expvAmain} is simply
  obtained by replacing $\sin^2(t_m\ww/2)$ by its average $1/2$, which
  corresponds to the first term in Eq.~\eqref{eq:b2periodic-1},
  \begin{align}
    \label{first-line}
    \eva{A}_{\text{pre}} 
    &=
    -2{\rm Re} \sandwich{\alpha}{S(0)A S(0)}{\alpha}
    \nonumber\\&
    =
    2g^2
    \sum_{n,p}\int_{-\infty}^{\infty}\!\!d\ww\,
    \frac{y_{np}(\ww)}{(\ww - n\WW)(\ww - p\WW)}\;.
  \end{align}
  In this case the system synchronizes for $t_m \gg T$ (and $t_m\ll
  g^{-1}$) into a periodic evolution with values $\eva{A}_{t_m} =
  \eva{A}_{\text{pre}}$, before further heating takes place on longer
  timescales. This is the analogue of prethermalization in a quenched
  system.

  \section{Statistical description of the prethermalized state}\label{sec:PGE}
  The
  condition $y_{nn}(n\WW)=0$ for the absence of linear absorption is
  equivalent to the absence of resonances in Eq.~\eqref{eq:S1}. Outside
  the Fermi golden rule regime, the constants of motion \eqref{scom}
  are thus well-defined, and one can ask whether the prethermalized
  state can be described by a 
  Gibbs ensemble
    $\rho_{\tilde G} = \sum_{\lambda} e^{-\mu_\lambda
    \tilde{\mathcal{I}}_\lambda}/Z_{\tilde G}$
 \footnote{The statistical enesemble
    $\rho_{\tilde G}$ is the periodic Gibbs ensemble of
    Ref.~\onlinecite{Lazarides_PRL14} evaluated at stroboscopic
    times.}, where the Lagrange
  multipliers $\mu_\lambda$ are determined by the constraint from the
  initial state, $\eva{ \tilde{\mathcal{I}}_\lambda}_{0} =
  \text{tr}\big[ \rho_{\tilde G} \tilde{\mathcal{I}}_{\lambda}]$,
  \beq
  \eva{A}_{\text{pre}} = \text{tr}\big[ \rho_{\tilde G} A].
  \eeq
  Using Eqs.~\eqref{first-line} and \eqref{scom}, the proof for this
  statement only relies on 
the time-independent matrix $S(0)$ being  antihermitian and appearing only to order $g$ in $\text{tr}\big[
  \rho_{\tilde G} A]$, and thus proceeds analogously to the argument showing that
  prethermalized states for a sudden quench can be described by a
  GGE~\cite{Kollar_PRB11}. 

  \section{Relation to the Floquet picture}\label{sec:floquet}
  We now 
  explain how
  the prethermalized state Eq.~\eqref{first-line} can be related to
  the Floquet spectrum of the Hamiltonian.  According to the Floquet
  theorem, the exact solution of the Schr\"odinger equation with a
  time-periodic Hamiltonian \eqref{eq:ham} is given in the form
  $\ket{\psi_{F,\alpha} (t)} = e^{-iE_{F,\alpha} t } \ket{\psi_\alpha
    (t)}$, where $\ket{\psi_\alpha (t)}$ is periodic in time. If a
  system is in a Floquet state, the time evolution of observables is
  periodic. By expanding $| \psi_\alpha (t) \rangle$ in a Fourier
  series $| \psi_\alpha (t) \rangle=\sum_{m} e^{-i\WW m t} |
  \psi_{\alpha,m} \rangle$, the Floquet quasi-energy spectrum can be
  obtained by diagonalizing the time-independent block-matrix,
  \begin{align}
    \label{floquet matrix}
    (E_{F,\alpha} + m\WW -H_0 ) | \psi_{\alpha,m} \rangle =  g \sum_{l} H_{{\rm int}, l} | \psi_{\alpha,m+l} \rangle.
  \end{align}
  In principle, one can now use standard first-order perturbation
  theory to construct perturbative Floquet states
  $\ket{\psi_{\alpha,n}}=\ket{\psi_{\alpha,n}^{(0)}}+g\ket{\psi_{\alpha,n}^{(1)}}+\cdots$,
  where the zeroth order is given by the unperturbed eigenstates
  $\ket{\psi_{\alpha,m}^{(0)}}=\delta_{m,0}\ket{\alpha}$
  ($E_{F,\alpha}^{(0)}=E_\alpha$). 
  The perturbative expansion does not converge to the true Floquet
  eigenstate if there are resonances $E_\alpha-E_\beta=n\WW$ in the 
  many-body spectrum, but low orders nevertheless can exist: in particular,
  the first order is given by $\ket{\psi_{\alpha,m}^{(1)}}=S_{m}^{(1)} \ket{\alpha}$,
  and it is well-defined outside the Fermi golden rule regime.   
  This shows
  that the prethermalized state Eq.~\eqref{first-line} is related to
  the perturbative Floquet state by
  \beq
  \eva{A}_{\text{pre}} =
  2
  \bra{\psi_{F,\alpha}^{(1)}}A\ket{\psi_{F,\alpha}^{(1)}}.
  \eeq
  Here the appearance of a factor of two is reminiscent to a similar
  relation between the prethermalized and ground state expectation
  values in the quench case.

  \section{Application to the Hubbard model 
  in infinite dimensions 
  }\label{sec:hubbard}
  \subsection{Analytical results}\label{ssec:analytical}
  In order to illustrate
  the general results above, we now choose as specific example the
  Hubbard model
  \beq\label{eq:hubb}
  H(t)=-J\sum_{\langle ij \rangle\sigma}c^{\dagger}_{i\sigma}c_{j\sigma}+
  U(t)\sum_{i}(n_{i\uparrow}-\tfrac12)(n_{i\downarrow}-\tfrac12)\,,
  \eeq
  with nearest neighbor hopping $J$ and periodically modulated
  interaction 
  \beq\label{eq:driving}
  U(t) = U(1-\cos(\Omega t))\,.
  \eeq
  With these choices, the
  first and the second term of Eq.~\eqref{eq:hubb} represent the
  integrable,
  noninteracting 
  part $H_0$ and the periodic 
  weakly interacting 
  perturbation with $g=U$, respectively.  Energy and time are measured
  in units of $J$ and $J^{-1}$, respectively.  The constants of motion
  of $H_0$ are momentum occupation numbers $\hat
  n_{\BK\sigma}=c^\dagger_{\BK\sigma} c_{\BK\sigma}$. To allow for a
  comparison of the analytical results derived above and a numerical
  solution, we consider the model in the limit of infinite spatial dimensions with a
  semi-elliptic density of states
  $\rho(\epsilon)$ $=$ $\sqrt{4-\epsilon^2}/(2\pi)$ at half-filling (density
  $n$ $=$ 1). In this limit, the dynamics can be computed using
  nonequilibrium dynamical-mean-field theory~\cite{REVIEW}, and
  iterative perturbation theory~\cite{Eckstein_PRB10,
    Tsuji2013weakcouplingprb} as impurity solver (see Sec.~\ref{ssec:numerical}).

  To investigate the prethermalization dynamics we use the momentum
  occupations as observables, 
  $A(t)\equiv n(\epsilon_\BK,t)-n(\epsilon_\BK,0)$,
  where $n(\epsilon_{\BK},0)$ is the initial occupation of the single-particle state
  with energy $\epsilon_{\BK}$. 
  For 
  the 
  harmonic
  driving 
  in Eq.~\eqref{eq:driving} 
  we have $H_{{\rm int
    },n}=h_nH_{{\rm int }}$ with $H_{{\rm int }}=\sum_i
  (n_{i\uparrow}-\tfrac12)(n_{i\downarrow}-\tfrac12)$, and $h_0=1$,
  $h_{\pm1}=-\tfrac12$. 
  We now proceed as in
  Refs.~\onlinecite{Eckstein_PRB10} and~\onlinecite{Kollar_PRB11},
  noting that in those derivations also initial free thermal
  states are allowed by virtue of the finite-temperature version of
  Wick's theorem.   
  The
  time-dependent occupation of a state with single-particle energy
  $\epsilon$ at time $t_m=mT$ is given by~\cite{Eckstein_PRB10,Kollar_PRB11}:
  \begin{align}\label{eq:npert}
  n_{\rm pert}(\epsilon_{\BK},t_m)=n(\epsilon_\BK,0)-4U^2F(\epsilon_\BK,t_m)\;,
  \end{align}
  where 
  we choose the initial distribution 
  $n(\epsilon_\BK,0)$ $=$ $\langle c^{\dagger}_{\BK\sigma}c_{\BK\sigma}\rangle_{t=0}$
  to be thermal,
  and
  \begin{align}
    F(\epsilon,t_m)\equiv&\sum_{n,p}
    \int\limits_{-\infty}^{\infty} d\ww\, \frac{\sin^2(\ww t_m/2)}{(\ww-n\WW)(\ww-p\WW)}h_n h_{-p} J_{\epsilon}(\omega)
    \nonumber\\
    \equiv&\sum_{n,p}h_nh_{-p}F_{n,p}(\epsilon,t_m)\,,\label{eq:F}
  \end{align}
  where we have dropped the $\BK$-dependence since
  momentum conservation can be omitted in the
  limit of infinite dimensions (i.e. 
  one has $J_{\BK}(\ww)$ $=$ $J_{\epsilon_\BK}(\ww)$). 
  We find:
  \begin{multline}\label{eq:J}
    \!\!\!\!\!\!
    J_{\epsilon}(\ww)
    =\int d\epsilon_1 d \epsilon_2  d\epsilon_3\, \rho(\epsilon_1)\rho(\epsilon_2)\rho(\epsilon_3)
    [n(\epsilon_3)n(\epsilon)\bar{n}(\epsilon_1)\bar{n}(\epsilon_2)
    \\
    -n(\epsilon_1)n(\epsilon_2)\bar{n}(\epsilon_3)\bar{n}(\epsilon)]
    \delta(\epsilon_1+\epsilon_2-\epsilon_3-\ww-\epsilon)\,,
  \end{multline}
  where $\bar{n}(\epsilon)\equiv 1-n(\epsilon)$ (which equals
  $n(-\epsilon)$ in the case of particle-hole symmetry, which we
 consider here).
 The function $J_{\epsilon}(\ww)$ (Eq.~\eqref{eq:J}) has already
  been obtained for the investigation of the sudden
  quench~\cite{Kollar_PRB11} (which is contained in our results by
  setting $h_{\pm1}=0)$. The connection with the spectral density~\eqref{eq:yfunction-1} is given by $y_{\epsilon, np}(\ww ) =
  -h_{n}h_{p}J_{\epsilon}(\ww)$.

  From Eq.~\eqref{eq:J}, one can read
  off the phase space condition for 
  the
  Fermi golden rule: at zero
  temperature, $n(\epsilon)$ $=$ $\Theta(-\epsilon)$ and $\rho(\epsilon)$ $=$ $0$
  for $|\epsilon|>2$, hence linear absorption ($J_{\BK}(\pm\WW)\neq
  0$) should occur for for $|\epsilon_{\BK}|<\WW<6+|\epsilon_{\BK}|$.
  More details on the phase-space argument leading to either the Fermi-golden rule
  regime or the prethermalization plateau can be found in the Appendix, where useful expressions for the numerical evaluation of Eq.~\eqref{eq:J} are also presented.

\subsection{Numerical results}\label{ssec:numerical}
%
  
  \begin{figure}[tb]
    \includegraphics[angle=0,width=1.0\columnwidth]{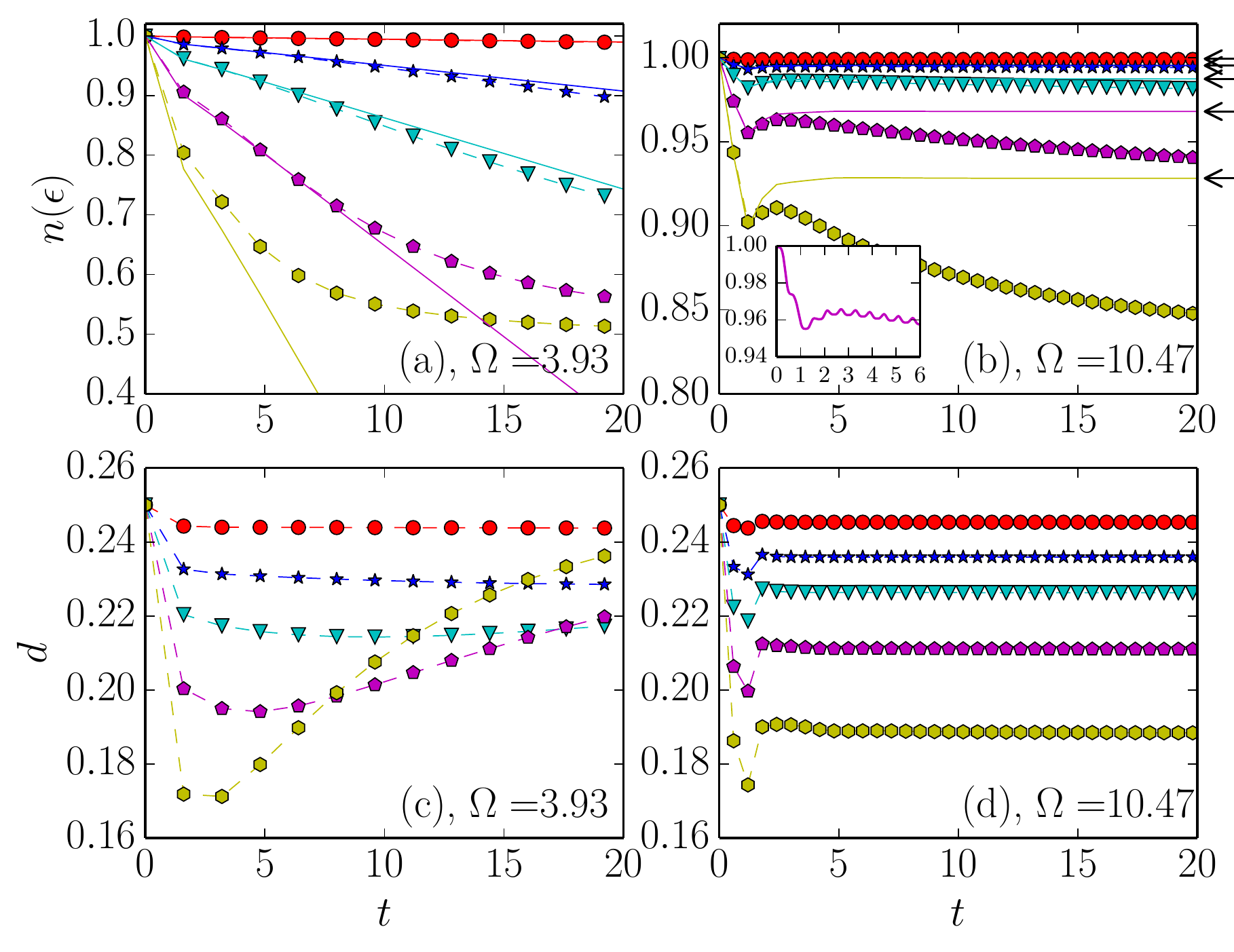}
    \caption{
    Momentum occupation $n(\epsilon_{\BK})$ for energy $\epsilon_{\BK}=-0.4$ 
    (upper panels) and double occupation (lower panels) 
    at stroboscopic times $2\pi m/\WW$ for $U=0.1,0.3,0.5, 0.8, 1.2$ from top to bottom curves in each panel. Panels (a) and (c): $\Omega=3.93$, panels (b) and (d): $\Omega=10.47$. Symbols with dashed lines: DMFT data, continuous lines: perturbative predictions from Eqs.~\eqref{eq:expvAmain} and~\eqref{eq:J}.
    Inset of
     panel (b): same as main panel for $U=0.8$, but showing complete time evolution (i.e. also non stroboscopic times). The arrows in 
     panel (b) show the predictions of Eq.~\eqref{first-line}.}
    
    \label{fig:recap}
  \end{figure}
  In Fig.~\ref{fig:recap} we show the single-particle occupation
  $n(\epsilon_{\BK})$ at stroboscopic times for a specific value of
  $\epsilon$ for 
  $\WW=3.93$ and $\WW=10.47$, 
  which
  lie in the Fermi golden-rule regime and in the prethermalization
  regime, respectively. We find that the perturbative predictions from
  Eqs.~\eqref{eq:expvAmain} and~\eqref{eq:J} 
  capture well the
  initial slope of the occupation in the linear absorption regime, as
  well as the prethermalization plateau 
  predicted by Eq.~\eqref{first-line} 
  for $\WW=10.47$. 
  For later
  times the numerical results approach the infinite-temperature value
  $n_{\beta\to 0}(\epsilon)$ $=$ $0.5$. As expected, the agreement between
  the DMFT results and the perturbative predictions improves with
  decreasing $U$, where the prethermalization plateau extends to
  longer times.
  In the inset of Fig.~\ref{fig:recap} we show the time evolution of
  the occupation $n(\epsilon)$ at $U=0.8$. At $t\sim2$-$4$ the
  quasi-periodic prethermalization regime begins where $n(\epsilon)$
  is constant at stroboscopic times.
    The double occupation 
  (lower panels of Fig.~\ref{fig:recap}) 
  also shows a prethermalization plateau at high frequency,
  while it 
  evolves towards its infinite temperature value for $\WW$ in the Fermi golden-rule regime.  
  \begin{figure}[!t]
    \includegraphics[angle=0,width=0.95\columnwidth]{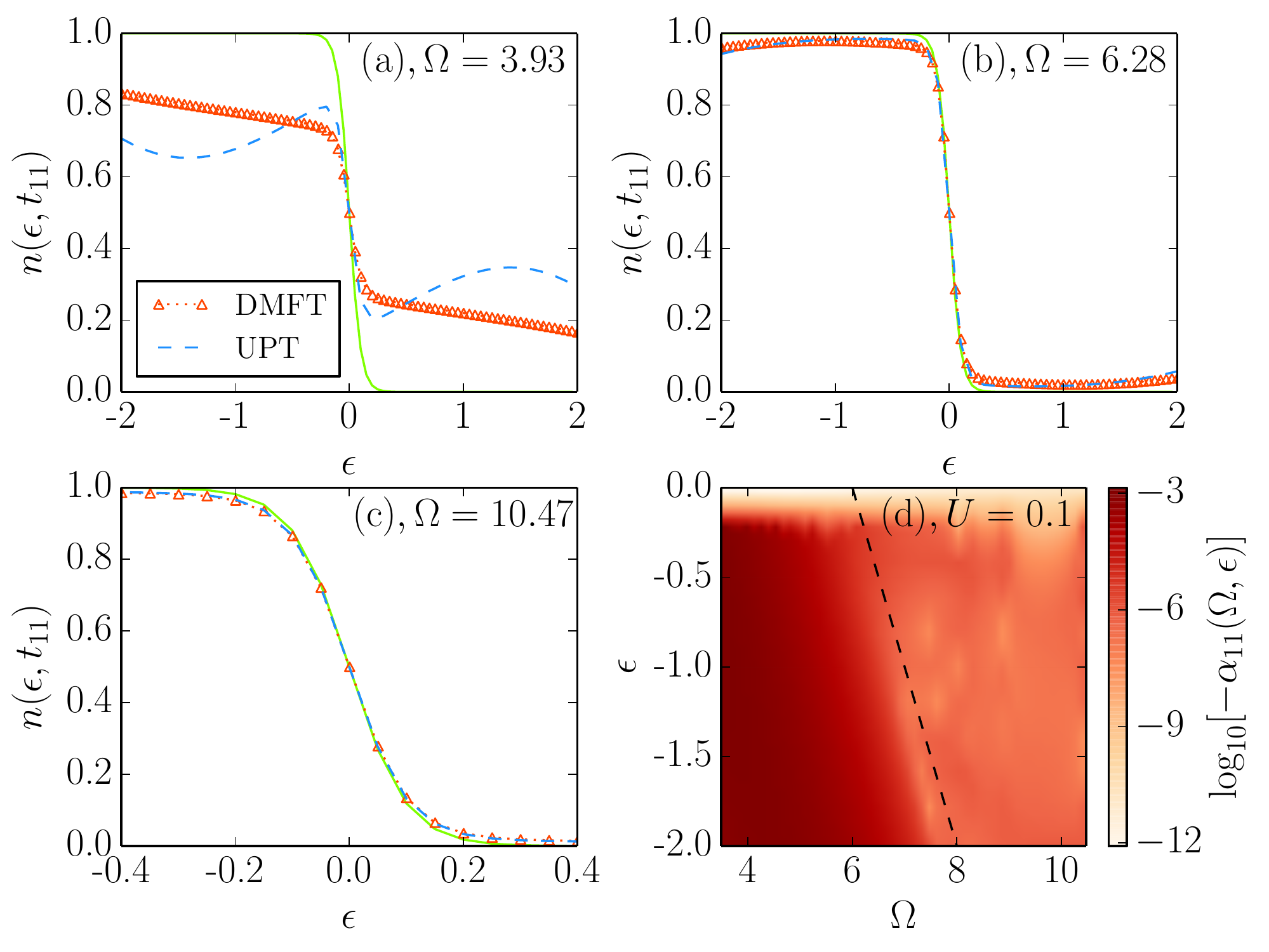}
    \caption{(a), (b), (c): single particle occupations for a driving
      term with $U=0.5$, computed with DMFT (red triangles, with
      initial inverse temperature $\beta$ $=$ $20$) and from
      perturbation theory (blue dashed lines, initially in the ground
      state) at an intermediate time, $t_{11}$ $=$ $11T$, for
      $\Omega$ $=$ $2\pi/T$ $=$ $3.93$, $6.28$, and $10.47$, respectively; the initial
      state is prepared with $U$ $=$ $0$.
	  The green continuous line shows the free
      initial thermal state at $\beta$ $=$ $20$. 
      The UPT prediction in panel (c) negligibly differs from the prethermalization plateau predicted by Eq.~\eqref{first-line}.
      (d): 
      Excitation over one period, measured
      as $\alpha_m(\epsilon,\WW)$ (see main text) for $U=0.1$ using DMFT. The black dashed line 
      corresponds to $\epsilon=-\WW+6$, which together with $\epsilon=-\WW$ delimits the Fermi-golden-rule regime.
    }
    \label{fig:eps}
  \end{figure}
  In Fig.~\ref{fig:eps} we plot $n(\epsilon,t_m)$ as a function of
  $\epsilon$ after a given number of periods ($m=11$). Panel (a)
  corresponds to a frequency such that every value of $\epsilon$ gives
  rise to linear terms, which are on the contrary absent for
  $\WW=10.47$, see panel (c). Panel (b) refers to an intermediate case
  ($T=1.0$, $\WW=2\pi$), where only the boundary values of $\epsilon$
  (i.e. $\epsilon\gtrsim -2$ and $\epsilon\lesssim 2$) give linear
  contributions and thus at 
  $t_m\gg T$ 
  differ from the DMFT data.
  Finally, panel (d) shows the 
  absorption of energy, measured by the
  slope $\alpha_m(\epsilon,\WW)\equiv
  n(\epsilon,m 2\pi/\WW)-n(\epsilon,(m-1)2\pi/\WW)$, 
  which becomes small for $\Omega> 6-\epsilon$, as predicted by the
  perturbative calculation (shown with a dashed line for $m$ $=$
  $11$). 
  We point out that the regime of validity of the DMFT
  calculation with iterative perturbation theory does not allow to
  explore small values of the frequency ($\Omega\lesssim 1$) where the
  other boundary ($\epsilon=-\Omega$) lies.

  \section{Conclusions}\label{sec:conclusions}
  In conclusion, we discussed the analogue of
  prethermalization in periodically driven systems. A weakly
  interacting system can synchronize into a quasi-steady state with
  nontrivial properties, before reaching the infinite temperature
  state generic for the long-time behavior of driven non-integrable
  systems. This stroboscopic prethermalization is a consequence of the
  existence of a macroscopic set of operators which are almost
  conserved by the time evolution over one period. 
Stroboscopic prethermalization thus provides a way to engineer 
quantum states with a nontrivial effective dynamics, alternative to the 
a high frequency expansion. These states reflect the properties of 
perturbative Floquet states, which can be very different in nature from the exact Floquet states.

  \begin{acknowledgments}
    The authors would like to acknowledge fruitful discussions with
    Markus Heyl and Hugo U.\ R.\ Strand.  M.K.\ was 
    supported  
    in part by
    Transregio 80 of Deutsche Forschungsgemeinschaft.
  \end{acknowledgments}

\appendix
\section{Phase-space argument for the Hubbard model in infinite-dimensions}
As discussed in Sec.~\ref{ssec:obs}, the single-particle occupations
  Eq.~\eqref{eq:npert} at long times (i.e. $t_m\gg T$) display two
  regimes, namely the Fermi-golden rule absorption regime and the
  stroboscopic prethermalization regime, depending on the value of
  $\Omega$ and $\epsilon$. Here we discuss these regimes for the
  specific case of the driven Hubbard interaction by rewriting
  Eq.~\eqref{eq:F} and applying a phase-space argument.

  As a first step, we express $J_{\epsilon}(\ww)$ in terms of
  \begin{align}
  R(s)
  &\equiv
  \int\limits_{-\infty}^{\infty} d\epsilon\,
  n(\epsilon)\,
  \rho(\epsilon)\,
  e^{is\epsilon}\,,
  \end{align}
  using a Fourier representation of the delta function:
  \begin{align}
    J_{\epsilon}(\ww)
    &=
    \int\limits_{-\infty}^{\infty}\frac{ds}{2\pi}
    \left[
      n(\epsilon)e^{i(\epsilon+\omega)s}
      -
      \bar{n}(\epsilon)e^{-i(\epsilon+\omega)s}
    \right]R(s)^3
    \,.
  \end{align}
  We also note that for an initial zero-temperature state (with
  $n(\epsilon)$ $=$ $\Theta(-\epsilon)$), $J_{\epsilon}(\ww)$ is zero
  unless $|\epsilon|$ $\leq$ $|\omega|$ $\leq$ $3D+|\epsilon|$, where
  $D$ is the half-bandwidth.

  A partial fraction decomposition of the functions in
  Eq.~\eqref{eq:F} and a shift of the integration variable yield
  \begin{align}
    F_{n,p}(\epsilon,t_m)
    &=
    \frac{F^{(1)}(\epsilon,t_m,n\Omega)-F^{(1)}(\epsilon,t_m,p\Omega)}{(n-p)\Omega}\,,
    ~
    (n\neq p)
    \nonumber\\
    F_{n,n}(\epsilon,t_m)
    &=
    F^{(2)}(\epsilon,t_m,n\Omega)\,,
  \end{align}
  where we defined
  \begin{align}
    F^{(N)}(\epsilon,t_m,E)
    &\equiv
    \int\limits_{-\infty}^{\infty} d\ww\,
    \frac{\sin^2(\ww t_m/2)}{\ww^N}J_{\epsilon}(\omega+E)
    \,.
  \end{align}
  Consider first the case of zero (or sufficiently low) temperature of
  the initial state and $|\epsilon|$ $\leq$ $|\WW|$ $\leq$
  $3D+|\epsilon|$. Then a term linear in $t_m$ contributes to
  $F(\epsilon,t_m)$, namely ($E$ $=$ $|n\WW|$, $N$ $=$ 1,2, $x$ $=$
  $(\ww-E)t_m$)
  \begin{align}
    F^{(N)}(\epsilon,t_m,E)
    &=
    t_m^{N-1}\int\limits_{-\infty}^{\infty} dx\,
    \frac{\sin^2(x/2)}{x^N}J_{\epsilon}(\tfrac{x}{t_m}+E)
    \nonumber\\&
    \sim
    \delta_{N2}
    \frac{\pi t_m}{2}
    J_{\epsilon}(E)
    \,~~(t_m\to\infty)
  \end{align}
  This corresponds to the Fermi golden rule regime with a
  linear-in-time growth of $n(\epsilon,t_m)$.
  On the other hand, if $\Omega$ is outside the indicated interval, the denominators are
  never zero (for zero temperature) and a stroboscopic
  prethermalization plateau is attained.
  
  In all cases we can rewrite the integrals more compactly by using the identities
  \begin{align}
    \frac{\sin^2(\ww t/2)}{\ww}
    &=
    \frac{1}{2}\int\limits_0^{t} du\,
    \sin(\omega u)
    \,,
    \\
    \int\limits_0^{\infty} d\ww\,
    \frac{\sin^2(\ww t/2)}{\ww^2}\cos(\ww s)
    &=\frac{\pi}{4}(t-s)\Theta(t-s)
    \,,
  \end{align}
  and taking the symmetries of the $\omega$ and $s$ integrals into
  account. We obtain  
  \begin{align}
    F^{(1)}(\epsilon,t_m,E)
    &=
    -\frac12
    \int\limits_{0}^{t_m} ds\,
    \text{Im}\bigg[
      R(s)^3\;\times
      \\\nonumber&~~~~~~
      \Big(n(\epsilon)e^{i(\epsilon+E)s}
      +
      \bar{n}(\epsilon)e^{-i(\epsilon+E)s}
      \Big)
      \bigg],
    \\
    F^{(2)}(\epsilon,t_m,E)
    &=
    \frac12
    \int\limits_{0}^{t_m} ds\,
    \text{Re}\bigg[
      R(s)^3\;\times
      \\\nonumber&~~~~~~
      \Big(n(\epsilon)e^{i(\epsilon+E)s}
      -
      \bar{n}(\epsilon)e^{-i(\epsilon+E)s}
      \Big)
      \bigg].      
  \end{align}
  These expressions are suitable for numerical evaluation; they can be
  further simplified for the zero-temperature case.

  \bibliography{rmpbib}

\begin{thebibliography}{46}%
\makeatletter
\providecommand \@ifxundefined [1]{%
 \@ifx{#1\undefined}
}%
\providecommand \@ifnum [1]{%
 \ifnum #1\expandafter \@firstoftwo
 \else \expandafter \@secondoftwo
 \fi
}%
\providecommand \@ifx [1]{%
 \ifx #1\expandafter \@firstoftwo
 \else \expandafter \@secondoftwo
 \fi
}%
\providecommand \natexlab [1]{#1}%
\providecommand \enquote  [1]{``#1''}%
\providecommand \bibnamefont  [1]{#1}%
\providecommand \bibfnamefont [1]{#1}%
\providecommand \citenamefont [1]{#1}%
\providecommand \href@noop [0]{\@secondoftwo}%
\providecommand \href [0]{\begingroup \@sanitize@url \@href}%
\providecommand \@href[1]{\@@startlink{#1}\@@href}%
\providecommand \@@href[1]{\endgroup#1\@@endlink}%
\providecommand \@sanitize@url [0]{\catcode `\\12\catcode `\$12\catcode
  `\&12\catcode `\#12\catcode `\^12\catcode `\_12\catcode `\%12\relax}%
\providecommand \@@startlink[1]{}%
\providecommand \@@endlink[0]{}%
\providecommand \url  [0]{\begingroup\@sanitize@url \@url }%
\providecommand \@url [1]{\endgroup\@href {#1}{\urlprefix }}%
\providecommand \urlprefix  [0]{URL }%
\providecommand \Eprint [0]{\href }%
\providecommand \doibase [0]{http://dx.doi.org/}%
\providecommand \selectlanguage [0]{\@gobble}%
\providecommand \bibinfo  [0]{\@secondoftwo}%
\providecommand \bibfield  [0]{\@secondoftwo}%
\providecommand \translation [1]{[#1]}%
\providecommand \BibitemOpen [0]{}%
\providecommand \bibitemStop [0]{}%
\providecommand \bibitemNoStop [0]{.\EOS\space}%
\providecommand \EOS [0]{\spacefactor3000\relax}%
\providecommand \BibitemShut  [1]{\csname bibitem#1\endcsname}%
\let\auto@bib@innerbib\@empty
\bibitem [{\citenamefont {Shirley}(1965)}]{Shirley1965}%
  \BibitemOpen
  \bibfield  {author} {\bibinfo {author} {\bibfnamefont {J.~H.}\ \bibnamefont
  {Shirley}},\ }\href@noop {} {\bibfield  {journal} {\bibinfo  {journal} {Phys.
  Rev.}\ }\textbf {\bibinfo {volume} {138}},\ \bibinfo {pages} {B979} (\bibinfo
  {year} {1965})}\BibitemShut {NoStop}%
\bibitem [{\citenamefont {Sambe}(1973)}]{Sambe1973}%
  \BibitemOpen
  \bibfield  {author} {\bibinfo {author} {\bibfnamefont {H.}~\bibnamefont
  {Sambe}},\ }\href@noop {} {\bibfield  {journal} {\bibinfo  {journal} {Phys.
  Rev. A}\ }\textbf {\bibinfo {volume} {7}},\ \bibinfo {pages} {2203} (\bibinfo
  {year} {1973})}\BibitemShut {NoStop}%
\bibitem [{\citenamefont {{Breuer, H.P.}}, \citenamefont {{Dietz, K.}},\ and\
  \citenamefont {{Holthaus, M.}}(1990)}]{Breuer_JPF90}%
  \BibitemOpen
  \bibfield  {author} {\bibinfo {author} {\bibnamefont {{Breuer, H.P.}}},
  \bibinfo {author} {\bibnamefont {{Dietz, K.}}}, \ and\ \bibinfo {author}
  {\bibnamefont {{Holthaus, M.}}},\ }\href {\doibase
  10.1051/jphys:01990005108070900} {\bibfield  {journal} {\bibinfo  {journal}
  {J. Phys. France}\ }\textbf {\bibinfo {volume} {51}},\ \bibinfo {pages} {709}
  (\bibinfo {year} {1990})}\BibitemShut {NoStop}%
\bibitem [{\citenamefont {Grifoni}\ and\ \citenamefont
  {H\"{a}nggi}(1998)}]{GrifoniHanggi1998}%
  \BibitemOpen
  \bibfield  {author} {\bibinfo {author} {\bibfnamefont {M.}~\bibnamefont
  {Grifoni}}\ and\ \bibinfo {author} {\bibfnamefont {P.}~\bibnamefont
  {H\"{a}nggi}},\ }\href@noop {} {\bibfield  {journal} {\bibinfo  {journal}
  {Phys. Rep.}\ }\textbf {\bibinfo {volume} {304}},\ \bibinfo {pages} {229}
  (\bibinfo {year} {1998})}\BibitemShut {NoStop}%
\bibitem [{\citenamefont {Oka}\ and\ \citenamefont {Aoki}(2009)}]{Oka_PRB09}%
  \BibitemOpen
  \bibfield  {author} {\bibinfo {author} {\bibfnamefont {T.}~\bibnamefont
  {Oka}}\ and\ \bibinfo {author} {\bibfnamefont {H.}~\bibnamefont {Aoki}},\
  }\href {\doibase 10.1103/PhysRevB.79.081406} {\bibfield  {journal} {\bibinfo
  {journal} {Phys. Rev. B}\ }\textbf {\bibinfo {volume} {79}},\ \bibinfo
  {pages} {081406} (\bibinfo {year} {2009})}\BibitemShut {NoStop}%
\bibitem [{\citenamefont {Kitagawa}\ \emph {et~al.}(2011)\citenamefont
  {Kitagawa}, \citenamefont {Oka}, \citenamefont {Brataas}, \citenamefont
  {Fu},\ and\ \citenamefont {Demler}}]{Kitagawa_PRB11}%
  \BibitemOpen
  \bibfield  {author} {\bibinfo {author} {\bibfnamefont {T.}~\bibnamefont
  {Kitagawa}}, \bibinfo {author} {\bibfnamefont {T.}~\bibnamefont {Oka}},
  \bibinfo {author} {\bibfnamefont {A.}~\bibnamefont {Brataas}}, \bibinfo
  {author} {\bibfnamefont {L.}~\bibnamefont {Fu}}, \ and\ \bibinfo {author}
  {\bibfnamefont {E.}~\bibnamefont {Demler}},\ }\href {\doibase
  10.1103/PhysRevB.84.235108} {\bibfield  {journal} {\bibinfo  {journal} {Phys.
  Rev. B}\ }\textbf {\bibinfo {volume} {84}},\ \bibinfo {pages} {235108}
  (\bibinfo {year} {2011})}\BibitemShut {NoStop}%
\bibitem [{\citenamefont {Iadecola}\ \emph {et~al.}(2013)\citenamefont
  {Iadecola}, \citenamefont {Campbell}, \citenamefont {Chamon}, \citenamefont
  {Hou}, \citenamefont {Jackiw}, \citenamefont {Pi},\ and\ \citenamefont
  {Kusminskiy}}]{Iadecola_PRL13}%
  \BibitemOpen
  \bibfield  {author} {\bibinfo {author} {\bibfnamefont {T.}~\bibnamefont
  {Iadecola}}, \bibinfo {author} {\bibfnamefont {D.}~\bibnamefont {Campbell}},
  \bibinfo {author} {\bibfnamefont {C.}~\bibnamefont {Chamon}}, \bibinfo
  {author} {\bibfnamefont {C.-Y.}\ \bibnamefont {Hou}}, \bibinfo {author}
  {\bibfnamefont {R.}~\bibnamefont {Jackiw}}, \bibinfo {author} {\bibfnamefont
  {S.-Y.}\ \bibnamefont {Pi}}, \ and\ \bibinfo {author} {\bibfnamefont {S.~V.}\
  \bibnamefont {Kusminskiy}},\ }\href {\doibase 10.1103/PhysRevLett.110.176603}
  {\bibfield  {journal} {\bibinfo  {journal} {Phys. Rev. Lett.}\ }\textbf
  {\bibinfo {volume} {110}},\ \bibinfo {pages} {176603} (\bibinfo {year}
  {2013})}\BibitemShut {NoStop}%
\bibitem [{\citenamefont {Kitagawa}\ \emph {et~al.}(2010)\citenamefont
  {Kitagawa}, \citenamefont {Berg}, \citenamefont {Rudner},\ and\ \citenamefont
  {Demler}}]{KitagawaPRB10}%
  \BibitemOpen
  \bibfield  {author} {\bibinfo {author} {\bibfnamefont {T.}~\bibnamefont
  {Kitagawa}}, \bibinfo {author} {\bibfnamefont {E.}~\bibnamefont {Berg}},
  \bibinfo {author} {\bibfnamefont {M.}~\bibnamefont {Rudner}}, \ and\ \bibinfo
  {author} {\bibfnamefont {E.}~\bibnamefont {Demler}},\ }\href@noop {}
  {\bibfield  {journal} {\bibinfo  {journal} {Phys. Rev. B}\ }\textbf {\bibinfo
  {volume} {82}},\ \bibinfo {pages} {235114} (\bibinfo {year}
  {2010})}\BibitemShut {NoStop}%
\bibitem [{\citenamefont {Lindner}, \citenamefont {Refael},\ and\ \citenamefont
  {Galitski}(2011)}]{Lindner_NP11}%
  \BibitemOpen
  \bibfield  {author} {\bibinfo {author} {\bibfnamefont {N.}~\bibnamefont
  {Lindner}}, \bibinfo {author} {\bibfnamefont {G.}~\bibnamefont {Refael}}, \
  and\ \bibinfo {author} {\bibfnamefont {V.}~\bibnamefont {Galitski}},\ }\href
  {\doibase 10.1038/nphys1926} {\bibfield  {journal} {\bibinfo  {journal}
  {Nature Physics}\ }\textbf {\bibinfo {volume} {7}},\ \bibinfo {pages} {490}
  (\bibinfo {year} {2011})}\BibitemShut {NoStop}%
\bibitem [{\citenamefont {Wang}\ \emph {et~al.}(2013)\citenamefont {Wang},
  \citenamefont {Steinberg}, \citenamefont {Jarillo-Herrero},\ and\
  \citenamefont {Gedik}}]{Wang_SCI13}%
  \BibitemOpen
  \bibfield  {author} {\bibinfo {author} {\bibfnamefont {Y.~H.}\ \bibnamefont
  {Wang}}, \bibinfo {author} {\bibfnamefont {H.}~\bibnamefont {Steinberg}},
  \bibinfo {author} {\bibfnamefont {P.}~\bibnamefont {Jarillo-Herrero}}, \ and\
  \bibinfo {author} {\bibfnamefont {N.}~\bibnamefont {Gedik}},\ }\href
  {\doibase 10.1126/science.1239834} {\bibfield  {journal} {\bibinfo  {journal}
  {Science}\ }\textbf {\bibinfo {volume} {342}},\ \bibinfo {pages} {453}
  (\bibinfo {year} {2013})}\BibitemShut {NoStop}%
\bibitem [{\citenamefont {Fausti}\ \emph {et~al.}(2011)\citenamefont {Fausti},
  \citenamefont {Tobey}, \citenamefont {Dean}, \citenamefont {Kaiser},
  \citenamefont {Dienst}, \citenamefont {Hoffmann}, \citenamefont {Pyon},
  \citenamefont {Takayama}, \citenamefont {Takagi},\ and\ \citenamefont
  {Cavalleri}}]{Fausti11}%
  \BibitemOpen
  \bibfield  {author} {\bibinfo {author} {\bibfnamefont {D.}~\bibnamefont
  {Fausti}}, \bibinfo {author} {\bibfnamefont {R.~I.}\ \bibnamefont {Tobey}},
  \bibinfo {author} {\bibfnamefont {N.}~\bibnamefont {Dean}}, \bibinfo {author}
  {\bibfnamefont {S.}~\bibnamefont {Kaiser}}, \bibinfo {author} {\bibfnamefont
  {A.}~\bibnamefont {Dienst}}, \bibinfo {author} {\bibfnamefont {M.~C.}\
  \bibnamefont {Hoffmann}}, \bibinfo {author} {\bibfnamefont {S.}~\bibnamefont
  {Pyon}}, \bibinfo {author} {\bibfnamefont {T.}~\bibnamefont {Takayama}},
  \bibinfo {author} {\bibfnamefont {H.}~\bibnamefont {Takagi}}, \ and\ \bibinfo
  {author} {\bibfnamefont {A.}~\bibnamefont {Cavalleri}},\ }\href@noop {}
  {\bibfield  {journal} {\bibinfo  {journal} {Science}\ }\textbf {\bibinfo
  {volume} {331}},\ \bibinfo {pages} {189} (\bibinfo {year}
  {2011})}\BibitemShut {NoStop}%
\bibitem [{\citenamefont {Hu}\ \emph {et~al.}(2014)\citenamefont {Hu},
  \citenamefont {Kaiser}, \citenamefont {Nicoletti}, \citenamefont {Hunt},
  \citenamefont {Gierz}, \citenamefont {Hoffmann}, \citenamefont {{Le Tacon}},
  \citenamefont {Loew}, \citenamefont {Keimer},\ and\ \citenamefont
  {Cavalleri}}]{Hu2014}%
  \BibitemOpen
  \bibfield  {author} {\bibinfo {author} {\bibfnamefont {W.}~\bibnamefont
  {Hu}}, \bibinfo {author} {\bibfnamefont {S.}~\bibnamefont {Kaiser}}, \bibinfo
  {author} {\bibfnamefont {D.}~\bibnamefont {Nicoletti}}, \bibinfo {author}
  {\bibfnamefont {C.~R.}\ \bibnamefont {Hunt}}, \bibinfo {author}
  {\bibfnamefont {I.}~\bibnamefont {Gierz}}, \bibinfo {author} {\bibfnamefont
  {M.~C.}\ \bibnamefont {Hoffmann}}, \bibinfo {author} {\bibfnamefont
  {M.}~\bibnamefont {{Le Tacon}}}, \bibinfo {author} {\bibfnamefont
  {T.}~\bibnamefont {Loew}}, \bibinfo {author} {\bibfnamefont {B.}~\bibnamefont
  {Keimer}}, \ and\ \bibinfo {author} {\bibfnamefont {A.}~\bibnamefont
  {Cavalleri}},\ }\href {http://dx.doi.org/10.1038/nmat3963 10.1038/nmat3963
  http://www.nature.com/nmat/journal/v13/n7/abs/nmat3963.html\#supplementary-information}
  {\bibfield  {journal} {\bibinfo  {journal} {Nat. Mater.}\ }\textbf {\bibinfo
  {volume} {13}},\ \bibinfo {pages} {705} (\bibinfo {year} {2014})}\BibitemShut
  {NoStop}%
\bibitem [{\citenamefont {Goldman}\ and\ \citenamefont
  {Dalibard}(2014)}]{Goldman_PRX14}%
  \BibitemOpen
  \bibfield  {author} {\bibinfo {author} {\bibfnamefont {N.}~\bibnamefont
  {Goldman}}\ and\ \bibinfo {author} {\bibfnamefont {J.}~\bibnamefont
  {Dalibard}},\ }\href {\doibase 10.1103/PhysRevX.4.031027} {\bibfield
  {journal} {\bibinfo  {journal} {Phys. Rev. X}\ }\textbf {\bibinfo {volume}
  {4}},\ \bibinfo {pages} {031027} (\bibinfo {year} {2014})}\BibitemShut
  {NoStop}%
\bibitem [{\citenamefont {Mentink}, \citenamefont {Balzer},\ and\ \citenamefont
  {Eckstein}(2015)}]{Mentink2015}%
  \BibitemOpen
  \bibfield  {author} {\bibinfo {author} {\bibfnamefont {J.~H.}\ \bibnamefont
  {Mentink}}, \bibinfo {author} {\bibfnamefont {K.}~\bibnamefont {Balzer}}, \
  and\ \bibinfo {author} {\bibfnamefont {M.}~\bibnamefont {Eckstein}},\ }\href
  {\doibase 10.1038/ncomms7708} {\bibfield  {journal} {\bibinfo  {journal}
  {Nat. Commun.}\ }\textbf {\bibinfo {volume} {6}},\ \bibinfo {pages} {6708}
  (\bibinfo {year} {2015})}\BibitemShut {NoStop}%
\bibitem [{\citenamefont {{Mikhaylovskiy}}\ \emph {et~al.}()\citenamefont
  {{Mikhaylovskiy}}, \citenamefont {{Hendry}}, \citenamefont {{Secchi}},
  \citenamefont {{Mentink}}, \citenamefont {{Eckstein}}, \citenamefont {{Wu}},
  \citenamefont {{Pisarev}}, \citenamefont {{Kruglyak}}, \citenamefont
  {{Katsnelson}}, \citenamefont {{Rasing}},\ and\ \citenamefont
  {{Kimel}}}]{Mikhaylovskiy2014}%
  \BibitemOpen
  \bibfield  {author} {\bibinfo {author} {\bibfnamefont {R.~V.}\ \bibnamefont
  {{Mikhaylovskiy}}}, \bibinfo {author} {\bibfnamefont {E.}~\bibnamefont
  {{Hendry}}}, \bibinfo {author} {\bibfnamefont {A.}~\bibnamefont {{Secchi}}},
  \bibinfo {author} {\bibfnamefont {J.~H.}\ \bibnamefont {{Mentink}}}, \bibinfo
  {author} {\bibfnamefont {M.}~\bibnamefont {{Eckstein}}}, \bibinfo {author}
  {\bibfnamefont {A.}~\bibnamefont {{Wu}}}, \bibinfo {author} {\bibfnamefont
  {R.~V.}\ \bibnamefont {{Pisarev}}}, \bibinfo {author} {\bibfnamefont {V.~V.}\
  \bibnamefont {{Kruglyak}}}, \bibinfo {author} {\bibfnamefont {M.~I.}\
  \bibnamefont {{Katsnelson}}}, \bibinfo {author} {\bibfnamefont
  {T.}~\bibnamefont {{Rasing}}}, \ and\ \bibinfo {author} {\bibfnamefont
  {A.~V.}\ \bibnamefont {{Kimel}}},\ }\href@noop {} {\ }\Eprint
  {http://arxiv.org/abs/1412.7094} {arXiv:1412.7094} \BibitemShut {NoStop}%
\bibitem [{\citenamefont {{Itin}}\ and\ \citenamefont
  {{Katsnelson}}()}]{Itin_pp14}%
  \BibitemOpen
  \bibfield  {author} {\bibinfo {author} {\bibfnamefont {A.~P.}\ \bibnamefont
  {{Itin}}}\ and\ \bibinfo {author} {\bibfnamefont {M.~I.}\ \bibnamefont
  {{Katsnelson}}},\ }\href@noop {} {\ }\Eprint {http://arxiv.org/abs/1401.0402}
  {arXiv:1401.0402} \BibitemShut {NoStop}%
\bibitem [{\citenamefont {F\"{o}rst}\ \emph {et~al.}(2011)\citenamefont
  {F\"{o}rst}, \citenamefont {Manzoni}, \citenamefont {Kaiser}, \citenamefont
  {Tomioka}, \citenamefont {Tokura}, \citenamefont {Merlin},\ and\
  \citenamefont {Cavalleri}}]{Foerst2011}%
  \BibitemOpen
  \bibfield  {author} {\bibinfo {author} {\bibfnamefont {M.}~\bibnamefont
  {F\"{o}rst}}, \bibinfo {author} {\bibfnamefont {C.}~\bibnamefont {Manzoni}},
  \bibinfo {author} {\bibfnamefont {S.}~\bibnamefont {Kaiser}}, \bibinfo
  {author} {\bibfnamefont {Y.}~\bibnamefont {Tomioka}}, \bibinfo {author}
  {\bibfnamefont {Y.}~\bibnamefont {Tokura}}, \bibinfo {author} {\bibfnamefont
  {R.}~\bibnamefont {Merlin}}, \ and\ \bibinfo {author} {\bibfnamefont
  {A.}~\bibnamefont {Cavalleri}},\ }\href {\doibase 10.1038/nphys2055}
  {\bibfield  {journal} {\bibinfo  {journal} {Nature Physics}\ }\textbf
  {\bibinfo {volume} {7}},\ \bibinfo {pages} {854} (\bibinfo {year}
  {2011})}\BibitemShut {NoStop}%
\bibitem [{\citenamefont {Polkovnikov}\ \emph {et~al.}(2011)\citenamefont
  {Polkovnikov}, \citenamefont {Sengupta}, \citenamefont {Silva},\ and\
  \citenamefont {Vengalattore}}]{Polkovnikov2011RMP}%
  \BibitemOpen
  \bibfield  {author} {\bibinfo {author} {\bibfnamefont {A.}~\bibnamefont
  {Polkovnikov}}, \bibinfo {author} {\bibfnamefont {K.}~\bibnamefont
  {Sengupta}}, \bibinfo {author} {\bibfnamefont {A.}~\bibnamefont {Silva}}, \
  and\ \bibinfo {author} {\bibfnamefont {M.}~\bibnamefont {Vengalattore}},\
  }\href@noop {} {\bibfield  {journal} {\bibinfo  {journal} {Rev. Mod. Phys.}\
  }\textbf {\bibinfo {volume} {83}},\ \bibinfo {pages} {863} (\bibinfo {year}
  {2011})}\BibitemShut {NoStop}%
\bibitem [{\citenamefont {Kinoshita}, \citenamefont {Wenger},\ and\
  \citenamefont {Weiss}(2006)}]{Kinoshita2006}%
  \BibitemOpen
  \bibfield  {author} {\bibinfo {author} {\bibfnamefont {T.}~\bibnamefont
  {Kinoshita}}, \bibinfo {author} {\bibfnamefont {T.}~\bibnamefont {Wenger}}, \
  and\ \bibinfo {author} {\bibfnamefont {D.~S.}\ \bibnamefont {Weiss}},\
  }\href@noop {} {\bibfield  {journal} {\bibinfo  {journal} {Nature}\ }\textbf
  {\bibinfo {volume} {440}},\ \bibinfo {pages} {900} (\bibinfo {year}
  {2006})}\BibitemShut {NoStop}%
\bibitem [{\citenamefont {Trotzky}\ \emph {et~al.}(2012)\citenamefont
  {Trotzky}, \citenamefont {Chen}, \citenamefont {Flesch}, \citenamefont
  {McCulloch}, \citenamefont {Schollw\"{o}ck}, \citenamefont {Eisert},\ and\
  \citenamefont {Bloch}}]{Trotzky2012}%
  \BibitemOpen
  \bibfield  {author} {\bibinfo {author} {\bibfnamefont {S.}~\bibnamefont
  {Trotzky}}, \bibinfo {author} {\bibfnamefont {Y.-a.}\ \bibnamefont {Chen}},
  \bibinfo {author} {\bibfnamefont {A.}~\bibnamefont {Flesch}}, \bibinfo
  {author} {\bibfnamefont {I.~P.}\ \bibnamefont {McCulloch}}, \bibinfo {author}
  {\bibfnamefont {U.}~\bibnamefont {Schollw\"{o}ck}}, \bibinfo {author}
  {\bibfnamefont {J.}~\bibnamefont {Eisert}}, \ and\ \bibinfo {author}
  {\bibfnamefont {I.}~\bibnamefont {Bloch}},\ }\href {\doibase
  10.1038/nphys2232} {\bibfield  {journal} {\bibinfo  {journal} {Nature
  Physics}\ }\textbf {\bibinfo {volume} {8}},\ \bibinfo {pages} {325} (\bibinfo
  {year} {2012})}\BibitemShut {NoStop}%
\bibitem [{\citenamefont {Altshuler}\ \emph {et~al.}(1997)\citenamefont
  {Altshuler}, \citenamefont {Gefen}, \citenamefont {Kamenev},\ and\
  \citenamefont {Levitov}}]{Altshuler_PRL97}%
  \BibitemOpen
  \bibfield  {author} {\bibinfo {author} {\bibfnamefont {B.~L.}\ \bibnamefont
  {Altshuler}}, \bibinfo {author} {\bibfnamefont {Y.}~\bibnamefont {Gefen}},
  \bibinfo {author} {\bibfnamefont {A.}~\bibnamefont {Kamenev}}, \ and\
  \bibinfo {author} {\bibfnamefont {L.~S.}\ \bibnamefont {Levitov}},\ }\href
  {\doibase 10.1103/PhysRevLett.78.2803} {\bibfield  {journal} {\bibinfo
  {journal} {Phys. Rev. Lett.}\ }\textbf {\bibinfo {volume} {78}},\ \bibinfo
  {pages} {2803} (\bibinfo {year} {1997})}\BibitemShut {NoStop}%
\bibitem [{\citenamefont {Gornyi}, \citenamefont {Mirlin},\ and\ \citenamefont
  {Polyakov}(2005)}]{Gornyi_PRL05}%
  \BibitemOpen
  \bibfield  {author} {\bibinfo {author} {\bibfnamefont {I.~V.}\ \bibnamefont
  {Gornyi}}, \bibinfo {author} {\bibfnamefont {A.~D.}\ \bibnamefont {Mirlin}},
  \ and\ \bibinfo {author} {\bibfnamefont {D.~G.}\ \bibnamefont {Polyakov}},\
  }\href {\doibase 10.1103/PhysRevLett.95.206603} {\bibfield  {journal}
  {\bibinfo  {journal} {Phys. Rev. Lett.}\ }\textbf {\bibinfo {volume} {95}},\
  \bibinfo {pages} {206603} (\bibinfo {year} {2005})}\BibitemShut {NoStop}%
\bibitem [{\citenamefont {Basko}, \citenamefont {Aleiner},\ and\ \citenamefont
  {Altshuler}(2006)}]{Basko_AP06}%
  \BibitemOpen
  \bibfield  {author} {\bibinfo {author} {\bibfnamefont {D.}~\bibnamefont
  {Basko}}, \bibinfo {author} {\bibfnamefont {I.}~\bibnamefont {Aleiner}}, \
  and\ \bibinfo {author} {\bibfnamefont {B.}~\bibnamefont {Altshuler}},\ }\href
  {\doibase http://dx.doi.org/10.1016/j.aop.2005.11.014} {\bibfield  {journal}
  {\bibinfo  {journal} {Annals of Physics}\ }\textbf {\bibinfo {volume}
  {321}},\ \bibinfo {pages} {1126 } (\bibinfo {year} {2006})}\BibitemShut
  {NoStop}%
\bibitem [{\citenamefont {Rigol}\ \emph {et~al.}(2007)\citenamefont {Rigol},
  \citenamefont {Dunjko}, \citenamefont {Yurovsky},\ and\ \citenamefont
  {Olshanii}}]{Rigol2007}%
  \BibitemOpen
  \bibfield  {author} {\bibinfo {author} {\bibfnamefont {M.}~\bibnamefont
  {Rigol}}, \bibinfo {author} {\bibfnamefont {V.}~\bibnamefont {Dunjko}},
  \bibinfo {author} {\bibfnamefont {V.}~\bibnamefont {Yurovsky}}, \ and\
  \bibinfo {author} {\bibfnamefont {M.}~\bibnamefont {Olshanii}},\ }\href@noop
  {} {\bibfield  {journal} {\bibinfo  {journal} {Phys. Rev. Lett.}\ }\textbf
  {\bibinfo {volume} {98}},\ \bibinfo {pages} {050405} (\bibinfo {year}
  {2007})}\BibitemShut {NoStop}%
\bibitem [{\citenamefont {Langen}\ \emph {et~al.}(2015)\citenamefont {Langen},
  \citenamefont {Erne}, \citenamefont {Geiger}, \citenamefont {Rauer},
  \citenamefont {Schweigler}, \citenamefont {Kuhnert}, \citenamefont
  {Rohringer}, \citenamefont {Mazets}, \citenamefont {Gasenzer},\ and\
  \citenamefont {Schmiedmayer}}]{Langen2014}%
  \BibitemOpen
  \bibfield  {author} {\bibinfo {author} {\bibfnamefont {T.}~\bibnamefont
  {Langen}}, \bibinfo {author} {\bibfnamefont {S.}~\bibnamefont {Erne}},
  \bibinfo {author} {\bibfnamefont {R.}~\bibnamefont {Geiger}}, \bibinfo
  {author} {\bibfnamefont {B.}~\bibnamefont {Rauer}}, \bibinfo {author}
  {\bibfnamefont {T.}~\bibnamefont {Schweigler}}, \bibinfo {author}
  {\bibfnamefont {M.}~\bibnamefont {Kuhnert}}, \bibinfo {author} {\bibfnamefont
  {W.}~\bibnamefont {Rohringer}}, \bibinfo {author} {\bibfnamefont {I.~E.}\
  \bibnamefont {Mazets}}, \bibinfo {author} {\bibfnamefont {T.}~\bibnamefont
  {Gasenzer}}, \ and\ \bibinfo {author} {\bibfnamefont {J.}~\bibnamefont
  {Schmiedmayer}},\ }\href {\doibase 10.1126/science.1257026} {\bibfield
  {journal} {\bibinfo  {journal} {Science}\ }\textbf {\bibinfo {volume}
  {348}},\ \bibinfo {pages} {207} (\bibinfo {year} {2015})}\BibitemShut
  {NoStop}%
\bibitem [{\citenamefont {Kollar}, \citenamefont {Wolf},\ and\ \citenamefont
  {Eckstein}(2011)}]{Kollar_PRB11}%
  \BibitemOpen
  \bibfield  {author} {\bibinfo {author} {\bibfnamefont {M.}~\bibnamefont
  {Kollar}}, \bibinfo {author} {\bibfnamefont {F.~A.}\ \bibnamefont {Wolf}}, \
  and\ \bibinfo {author} {\bibfnamefont {M.}~\bibnamefont {Eckstein}},\ }\href
  {\doibase 10.1103/PhysRevB.84.054304} {\bibfield  {journal} {\bibinfo
  {journal} {Phys. Rev. B}\ }\textbf {\bibinfo {volume} {84}},\ \bibinfo
  {pages} {054304} (\bibinfo {year} {2011})}\BibitemShut {NoStop}%
\bibitem [{\citenamefont {Berges}, \citenamefont {Bors\'anyi},\ and\
  \citenamefont {Wetterich}(2004)}]{Berges2004a}%
  \BibitemOpen
  \bibfield  {author} {\bibinfo {author} {\bibfnamefont {J.}~\bibnamefont
  {Berges}}, \bibinfo {author} {\bibfnamefont {S.}~\bibnamefont {Bors\'anyi}},
  \ and\ \bibinfo {author} {\bibfnamefont {C.}~\bibnamefont {Wetterich}},\
  }\href@noop {} {\bibfield  {journal} {\bibinfo  {journal} {Phys. Rev. Lett.}\
  }\textbf {\bibinfo {volume} {93}},\ \bibinfo {pages} {142002} (\bibinfo
  {year} {2004})}\BibitemShut {NoStop}%
\bibitem [{\citenamefont {Moeckel}\ and\ \citenamefont
  {Kehrein}(2008)}]{Moeckel2008a}%
  \BibitemOpen
  \bibfield  {author} {\bibinfo {author} {\bibfnamefont {M.}~\bibnamefont
  {Moeckel}}\ and\ \bibinfo {author} {\bibfnamefont {S.}~\bibnamefont
  {Kehrein}},\ }\href@noop {} {\bibfield  {journal} {\bibinfo  {journal} {Phys.
  Rev. Lett.}\ }\textbf {\bibinfo {volume} {100}},\ \bibinfo {pages} {175702}
  (\bibinfo {year} {2008})}\BibitemShut {NoStop}%
\bibitem [{\citenamefont {Gring}\ \emph {et~al.}(2012)\citenamefont {Gring},
  \citenamefont {Kuhnert}, \citenamefont {Langen}, \citenamefont {Kitagawa},
  \citenamefont {Rauer}, \citenamefont {Schreitl}, \citenamefont {Mazets},
  \citenamefont {Smith}, \citenamefont {Demler},\ and\ \citenamefont
  {Schmiedmayer}}]{Gring2012}%
  \BibitemOpen
  \bibfield  {author} {\bibinfo {author} {\bibfnamefont {M.}~\bibnamefont
  {Gring}}, \bibinfo {author} {\bibfnamefont {M.}~\bibnamefont {Kuhnert}},
  \bibinfo {author} {\bibfnamefont {T.}~\bibnamefont {Langen}}, \bibinfo
  {author} {\bibfnamefont {T.}~\bibnamefont {Kitagawa}}, \bibinfo {author}
  {\bibfnamefont {B.}~\bibnamefont {Rauer}}, \bibinfo {author} {\bibfnamefont
  {M.}~\bibnamefont {Schreitl}}, \bibinfo {author} {\bibfnamefont
  {I.}~\bibnamefont {Mazets}}, \bibinfo {author} {\bibfnamefont {D.~A.}\
  \bibnamefont {Smith}}, \bibinfo {author} {\bibfnamefont {E.}~\bibnamefont
  {Demler}}, \ and\ \bibinfo {author} {\bibfnamefont {J.}~\bibnamefont
  {Schmiedmayer}},\ }\href@noop {} {\bibfield  {journal} {\bibinfo  {journal}
  {Science}\ }\textbf {\bibinfo {volume} {337}},\ \bibinfo {pages} {1318}
  (\bibinfo {year} {2012})}\BibitemShut {NoStop}%
\bibitem [{\citenamefont {Russomanno}, \citenamefont {Silva},\ and\
  \citenamefont {Santoro}(2012)}]{Russomanno_PRL12}%
  \BibitemOpen
  \bibfield  {author} {\bibinfo {author} {\bibfnamefont {A.}~\bibnamefont
  {Russomanno}}, \bibinfo {author} {\bibfnamefont {A.}~\bibnamefont {Silva}}, \
  and\ \bibinfo {author} {\bibfnamefont {G.~E.}\ \bibnamefont {Santoro}},\
  }\href {\doibase 10.1103/PhysRevLett.109.257201} {\bibfield  {journal}
  {\bibinfo  {journal} {Phys. Rev. Lett.}\ }\textbf {\bibinfo {volume} {109}},\
  \bibinfo {pages} {257201} (\bibinfo {year} {2012})}\BibitemShut {NoStop}%
\bibitem [{\citenamefont {Lazarides}, \citenamefont {Das},\ and\ \citenamefont
  {Moessner}(2014{\natexlab{a}})}]{Lazarides_PRL14}%
  \BibitemOpen
  \bibfield  {author} {\bibinfo {author} {\bibfnamefont {A.}~\bibnamefont
  {Lazarides}}, \bibinfo {author} {\bibfnamefont {A.}~\bibnamefont {Das}}, \
  and\ \bibinfo {author} {\bibfnamefont {R.}~\bibnamefont {Moessner}},\ }\href
  {\doibase 10.1103/PhysRevLett.112.150401} {\bibfield  {journal} {\bibinfo
  {journal} {Phys. Rev. Lett.}\ }\textbf {\bibinfo {volume} {112}},\ \bibinfo
  {pages} {150401} (\bibinfo {year} {2014}{\natexlab{a}})}\BibitemShut
  {NoStop}%
\bibitem [{\citenamefont {Lazarides}, \citenamefont {Das},\ and\ \citenamefont
  {Moessner}(2015)}]{Lazarides_PRL15}%
  \BibitemOpen
  \bibfield  {author} {\bibinfo {author} {\bibfnamefont {A.}~\bibnamefont
  {Lazarides}}, \bibinfo {author} {\bibfnamefont {A.}~\bibnamefont {Das}}, \
  and\ \bibinfo {author} {\bibfnamefont {R.}~\bibnamefont {Moessner}},\ }\href
  {\doibase 10.1103/PhysRevLett.115.030402} {\bibfield  {journal} {\bibinfo
  {journal} {Phys. Rev. Lett.}\ }\textbf {\bibinfo {volume} {115}},\ \bibinfo
  {pages} {030402} (\bibinfo {year} {2015})}\BibitemShut {NoStop}%
\bibitem [{\citenamefont {{Abanin}}, \citenamefont {{De Roeck}},\ and\
  \citenamefont {{Huveneers}}(2014)}]{Abanin_pp14}%
  \BibitemOpen
  \bibfield  {author} {\bibinfo {author} {\bibfnamefont {D.}~\bibnamefont
  {{Abanin}}}, \bibinfo {author} {\bibfnamefont {W.}~\bibnamefont {{De
  Roeck}}}, \ and\ \bibinfo {author} {\bibfnamefont {F.}~\bibnamefont
  {{Huveneers}}},\ }\href@noop {} {\bibfield  {journal} {\bibinfo  {journal}
  {ArXiv e-prints}\ } (\bibinfo {year} {2014})},\ \Eprint
  {http://arxiv.org/abs/1412.4752} {arXiv:1412.4752 [cond-mat.dis-nn]}
  \BibitemShut {NoStop}%
\bibitem [{\citenamefont {D'Alessio}\ and\ \citenamefont
  {Polkovnikov}(2013)}]{DAlessio_AP13}%
  \BibitemOpen
  \bibfield  {author} {\bibinfo {author} {\bibfnamefont {L.}~\bibnamefont
  {D'Alessio}}\ and\ \bibinfo {author} {\bibfnamefont {A.}~\bibnamefont
  {Polkovnikov}},\ }\href {\doibase
  http://dx.doi.org/10.1016/j.aop.2013.02.011} {\bibfield  {journal} {\bibinfo
  {journal} {Annals of Physics}\ }\textbf {\bibinfo {volume} {333}},\ \bibinfo
  {pages} {19 } (\bibinfo {year} {2013})}\BibitemShut {NoStop}%
\bibitem [{\citenamefont {Ponte}\ \emph
  {et~al.}(2015{\natexlab{a}})\citenamefont {Ponte}, \citenamefont {Chandran},
  \citenamefont {Papi\'c},\ and\ \citenamefont {Abanin}}]{Ponte_AP15}%
  \BibitemOpen
  \bibfield  {author} {\bibinfo {author} {\bibfnamefont {P.}~\bibnamefont
  {Ponte}}, \bibinfo {author} {\bibfnamefont {A.}~\bibnamefont {Chandran}},
  \bibinfo {author} {\bibfnamefont {Z.}~\bibnamefont {Papi\'c}}, \ and\
  \bibinfo {author} {\bibfnamefont {D.~A.}\ \bibnamefont {Abanin}},\ }\href
  {\doibase http://dx.doi.org/10.1016/j.aop.2014.11.008} {\bibfield  {journal}
  {\bibinfo  {journal} {Annals of Physics}\ }\textbf {\bibinfo {volume}
  {353}},\ \bibinfo {pages} {196 } (\bibinfo {year}
  {2015}{\natexlab{a}})}\BibitemShut {NoStop}%
\bibitem [{\citenamefont {Ponte}\ \emph
  {et~al.}(2015{\natexlab{b}})\citenamefont {Ponte}, \citenamefont
  {Papi\ifmmode~\acute{c}\else \'{c}\fi{}}, \citenamefont {Huveneers},\ and\
  \citenamefont {Abanin}}]{Ponte_PRL15}%
  \BibitemOpen
  \bibfield  {author} {\bibinfo {author} {\bibfnamefont {P.}~\bibnamefont
  {Ponte}}, \bibinfo {author} {\bibfnamefont {Z.}~\bibnamefont
  {Papi\ifmmode~\acute{c}\else \'{c}\fi{}}}, \bibinfo {author} {\bibfnamefont
  {F.}~\bibnamefont {Huveneers}}, \ and\ \bibinfo {author} {\bibfnamefont
  {D.~A.}\ \bibnamefont {Abanin}},\ }\href {\doibase
  10.1103/PhysRevLett.114.140401} {\bibfield  {journal} {\bibinfo  {journal}
  {Phys. Rev. Lett.}\ }\textbf {\bibinfo {volume} {114}},\ \bibinfo {pages}
  {140401} (\bibinfo {year} {2015}{\natexlab{b}})}\BibitemShut {NoStop}%
\bibitem [{\citenamefont {Roy}\ and\ \citenamefont {Das}(2015)}]{Roy_PRB15}%
  \BibitemOpen
  \bibfield  {author} {\bibinfo {author} {\bibfnamefont {A.}~\bibnamefont
  {Roy}}\ and\ \bibinfo {author} {\bibfnamefont {A.}~\bibnamefont {Das}},\
  }\href {\doibase 10.1103/PhysRevB.91.121106} {\bibfield  {journal} {\bibinfo
  {journal} {Phys. Rev. B}\ }\textbf {\bibinfo {volume} {91}},\ \bibinfo
  {pages} {121106} (\bibinfo {year} {2015})}\BibitemShut {NoStop}%
\bibitem [{\citenamefont {Lazarides}, \citenamefont {Das},\ and\ \citenamefont
  {Moessner}(2014{\natexlab{b}})}]{Lazarides_PRE14}%
  \BibitemOpen
  \bibfield  {author} {\bibinfo {author} {\bibfnamefont {A.}~\bibnamefont
  {Lazarides}}, \bibinfo {author} {\bibfnamefont {A.}~\bibnamefont {Das}}, \
  and\ \bibinfo {author} {\bibfnamefont {R.}~\bibnamefont {Moessner}},\ }\href
  {\doibase 10.1103/PhysRevE.90.012110} {\bibfield  {journal} {\bibinfo
  {journal} {Phys. Rev. E}\ }\textbf {\bibinfo {volume} {90}},\ \bibinfo
  {pages} {012110} (\bibinfo {year} {2014}{\natexlab{b}})}\BibitemShut
  {NoStop}%
\bibitem [{\citenamefont {D'Alessio}\ and\ \citenamefont
  {Rigol}(2014)}]{DAlessio_PRX14}%
  \BibitemOpen
  \bibfield  {author} {\bibinfo {author} {\bibfnamefont {L.}~\bibnamefont
  {D'Alessio}}\ and\ \bibinfo {author} {\bibfnamefont {M.}~\bibnamefont
  {Rigol}},\ }\href {\doibase 10.1103/PhysRevX.4.041048} {\bibfield  {journal}
  {\bibinfo  {journal} {Phys. Rev. X}\ }\textbf {\bibinfo {volume} {4}},\
  \bibinfo {pages} {041048} (\bibinfo {year} {2014})}\BibitemShut {NoStop}%
\bibitem [{\citenamefont {{Eckardt}}\ and\ \citenamefont
  {{Anisimovas}}()}]{Eckardt_pp15}%
  \BibitemOpen
  \bibfield  {author} {\bibinfo {author} {\bibfnamefont {A.}~\bibnamefont
  {{Eckardt}}}\ and\ \bibinfo {author} {\bibfnamefont {E.}~\bibnamefont
  {{Anisimovas}}},\ }\href@noop {} {\ }\Eprint
  {http://arxiv.org/abs/1502.06477} {arXiv:1502.06477} \BibitemShut {NoStop}%
\bibitem [{\citenamefont {Dittrich}\ \emph {et~al.}(1998)\citenamefont
  {Dittrich}, \citenamefont {H\"{a}nggi}, \citenamefont {Ingold}, \citenamefont
  {Kramer}, \citenamefont {Sch\"{o}n},\ and\ \citenamefont
  {Zwerger}}]{Dittrich1998}%
  \BibitemOpen
  \bibfield  {author} {\bibinfo {author} {\bibfnamefont {T.}~\bibnamefont
  {Dittrich}}, \bibinfo {author} {\bibfnamefont {P.}~\bibnamefont
  {H\"{a}nggi}}, \bibinfo {author} {\bibfnamefont {G.~L.}\ \bibnamefont
  {Ingold}}, \bibinfo {author} {\bibfnamefont {B.}~\bibnamefont {Kramer}},
  \bibinfo {author} {\bibfnamefont {G.}~\bibnamefont {Sch\"{o}n}}, \ and\
  \bibinfo {author} {\bibfnamefont {W.}~\bibnamefont {Zwerger}},\ }\href@noop
  {} {\emph {\bibinfo {title} {Quantum Transport and Dissipation}}}\ (\bibinfo
  {publisher} {Wiley-VCH},\ \bibinfo {address} {Weinheim},\ \bibinfo {year}
  {1998})\BibitemShut {NoStop}%
\bibitem [{Note1()}]{Note1}%
  \BibitemOpen
  \bibinfo {note} {Unitary transformations of the periodic Hamiltonian $H(t)$
  are also used in Ref.~\protect \rev@citealp {Abanin_pp14}, but in that case
  with the aim of finding a many-body localized cycle Hamiltonian $e^{i
  H_{\protect \rm cycle}T}\equiv \protect \mathcal {T} e^{-i\DOTSI \intop
  \ilimits@ _{0}^t dt'\protect \tmspace +\thinmuskip {.1667em}H(t')}$, i.e. a
  Hamiltonian with a set of true integrals of motion.}\BibitemShut {Stop}%
\bibitem [{Note2()}]{Note2}%
  \BibitemOpen
  \bibinfo {note} {The statistical enesemble $\rho _{\protect \mathaccentV
  {tilde}07EG}$ is the periodic Gibbs ensemble of Ref.~\protect \rev@citealp
  {Lazarides_PRL14} evaluated at stroboscopic times.}\BibitemShut {Stop}%
\bibitem [{\citenamefont {Aoki}\ \emph {et~al.}(2014)\citenamefont {Aoki},
  \citenamefont {Tsuji}, \citenamefont {Eckstein}, \citenamefont {Kollar},
  \citenamefont {Oka},\ and\ \citenamefont {Werner}}]{REVIEW}%
  \BibitemOpen
  \bibfield  {author} {\bibinfo {author} {\bibfnamefont {H.}~\bibnamefont
  {Aoki}}, \bibinfo {author} {\bibfnamefont {N.}~\bibnamefont {Tsuji}},
  \bibinfo {author} {\bibfnamefont {M.}~\bibnamefont {Eckstein}}, \bibinfo
  {author} {\bibfnamefont {M.}~\bibnamefont {Kollar}}, \bibinfo {author}
  {\bibfnamefont {T.}~\bibnamefont {Oka}}, \ and\ \bibinfo {author}
  {\bibfnamefont {P.}~\bibnamefont {Werner}},\ }\href {\doibase
  10.1103/RevModPhys.86.779} {\bibfield  {journal} {\bibinfo  {journal} {Rev.
  Mod. Phys.}\ }\textbf {\bibinfo {volume} {86}},\ \bibinfo {pages} {779}
  (\bibinfo {year} {2014})}\BibitemShut {NoStop}%
\bibitem [{\citenamefont {Eckstein}, \citenamefont {Kollar},\ and\
  \citenamefont {Werner}(2010)}]{Eckstein_PRB10}%
  \BibitemOpen
  \bibfield  {author} {\bibinfo {author} {\bibfnamefont {M.}~\bibnamefont
  {Eckstein}}, \bibinfo {author} {\bibfnamefont {M.}~\bibnamefont {Kollar}}, \
  and\ \bibinfo {author} {\bibfnamefont {P.}~\bibnamefont {Werner}},\ }\href
  {\doibase 10.1103/PhysRevB.81.115131} {\bibfield  {journal} {\bibinfo
  {journal} {Phys. Rev. B}\ }\textbf {\bibinfo {volume} {81}},\ \bibinfo
  {pages} {115131} (\bibinfo {year} {2010})}\BibitemShut {NoStop}%
\bibitem [{\citenamefont {Tsuji}\ and\ \citenamefont
  {Werner}(2013)}]{Tsuji2013weakcouplingprb}%
  \BibitemOpen
  \bibfield  {author} {\bibinfo {author} {\bibfnamefont {N.}~\bibnamefont
  {Tsuji}}\ and\ \bibinfo {author} {\bibfnamefont {P.}~\bibnamefont {Werner}},\
  }\href {\doibase 10.1103/PhysRevB.88.165115} {\bibfield  {journal} {\bibinfo
  {journal} {Phys. Rev. B}\ }\textbf {\bibinfo {volume} {88}},\ \bibinfo
  {pages} {165115} (\bibinfo {year} {2013})}\BibitemShut {NoStop}%
\end{thebibliography}%

 \end{document}